\documentclass[aps,prl,twocolumn,superscriptaddress,showpacs,amsmath,amssymb,
amsfonts,balancelastpage]{revtex4-1}
\usepackage{amsmath}
\usepackage{amssymb}
\usepackage{amsfonts}
\usepackage[justification=raggedright]{caption}
\usepackage{subcaption}
\usepackage{graphicx}
\usepackage{epsfig}
\usepackage{hyperref}
\usepackage{color}
\usepackage{xcolor}
\hypersetup{
colorlinks=true,
urlcolor=blue,
citecolor=blue,
linkcolor=blue,}

\begin{document}
\title{Anisotropic superconductivity in La(O,F)BiSeS crystals revealed by field-angle dependent Andreev reflection spectroscopy}

\author{Mohammad Aslam}
\affiliation{Department of Physical Sciences,
Indian Institute of Science Education and Research Mohali,
Sector 81, S. A. S. Nagar, Manauli, PO: 140306, India}

\author{Sirshendu Gayen}
\affiliation{Department of Physical Sciences,
Indian Institute of Science Education and Research Mohali,
Sector 81, S. A. S. Nagar, Manauli, PO: 140306, India} 
\author{Avtar Singh}
\affiliation{Department of Physical Sciences,
Indian Institute of Science Education and Research Mohali,
Sector 81, S. A. S. Nagar, Manauli, PO: 140306, India}
\author{Masashi Tanaka}
\affiliation{MANA, National Institute for Materials Science,
Tsukuba, Ibaraki 305-0047, Japan}
\author{Takuma Yamaki}
\affiliation{MANA, National Institute for Materials Science,
Tsukuba, Ibaraki 305-0047, Japan}
\author{Yoshihiko Takano}
\affiliation{MANA, National Institute for Materials Science,
Tsukuba, Ibaraki 305-0047, Japan}
\author{Goutam Sheet}
\email {goutam@iisermohali.ac.in}
\affiliation{Department of Physical Sciences,
Indian Institute of Science Education and Research Mohali,
Sector 81, S. A. S. Nagar, Manauli, PO: 140306, India}
\date{\today}

\begin{abstract}
From field-angle dependent Andreev reflection spectroscopy on single crystals of La(O,F)BiSeS, which belongs to the recently discovered BiCh$_2$ (Ch = S, Se) based layered superconductors, we found that the superconductivity in La(O,F)BiSeS is highly anisotropic. We measured a superconducting energy gap of 0.61 meV for current injected along $c$-axis at 1.5 K. Detailed temperature and magnetic field dependent studies of the gap also reveal the presence of unconventional pairing in La(O,F)BiSeS. We show that the observed anisotropic superconducting properties can be attributed to the anisotropy in the superconducting order parameter with a complex symmetry in superconducting La(O,F)BiSeS.
\end{abstract}

\maketitle
One of the key structural features that is common to almost all high temperature superconductors is the existence of alternating superconducting layers separated by insulating spacer layers \cite{Muller, wu, hightc, chen, Satomi, YM, X, Yoshikazu}. The classic examples are the high T$_c$ cuprates \cite{Muller, wu, hightc, chen, Makoto} and the ferro pnictides \cite{Kamihara,Kami,Y,Z,A} where superconductivity occurs in the CuO$_2$ and the FeAs layers respectively. The nature of the spacer layer, in fact, plays a prominent role in deciding the important superconducting parameters. Therefore, by chemically tuning the spacer layer a large number of new high T$_c$ superconductors have been derived out of the known layered high T$_c$ superconductors \cite{Hiro}. In the context of the cuprate and the ferro pnictide superconductors it has also been observed that owing to their layered structures comprising of layers with varying electronic properties, the superconducting properties are highly anisotropic and the pairing mechanism is unconventional \cite{D,Ja,Ka}. It is believed that detailed understanding of the superconducting parameters in such systems will eventually lead to the discovery of superconductors with higher critical temperatures.

Recently, a new class of layered superconductors comprising of alternate stacks of BiCh$_2$ (Ch= S, Se) layers separated by a number of spacer layers have been discovered. Some of the members of this new family of superconductors display exceptional richness in their phase diagram where even apparently antagonistic phenomena like superconductivity and ferro magnetism coexist \cite{Li,Naka}. Transport measurements in presence of pressure reveal that the critical temperature of the BiS$_2$ based superconductors increase with pressure signifying the importance of the crystal structure for the occurrence of superconductivity \cite{C,G,Pall}. Based on such measurements and pressure dependent structural analysis it has been thought that superconductivity in this system might be governed by lattice expansion. However, the nature of superconductivity in this family of superconductors is an outstanding issue which must be addressed. In this paper from field-angle dependent point contact Andreev reflection (PCAR) spectroscopy\cite{Naidyuk}  measurements on single crystals of La(O,F)BiSeS we show that the pairing in these class of superconductors is unconventional and highly anisotropic. 

The point contact Andreev reflection spectroscopy measurements were carried out in a liquid helium based cryostat equipped with a 3-axis vector magnet (6T-1T-1T). A conventional needle-anvil method was used for forming ballistic point contacts on single crystals of La(O,F)BiSeS for energy resolved spectroscopy. The differential conductance $dI/dV$ vs. $V$ spectra were recorded by a lock-in modulation technique at 1.7 kHz. It should be noted that all the spectra presented here are free from critical current driven conductance dips and other contact-heating related artefacts.\cite{Goutam_PRB} Therefore, all our spectra can be used for extracting energy resolved spectroscopic information. The ballistic nature of the presented point contacts have been confirmed by the measurement of the normal state resistance which remained temperature-independent.

\begin{figure}[h!]
	\centering
		\includegraphics[width=0.4\textwidth]{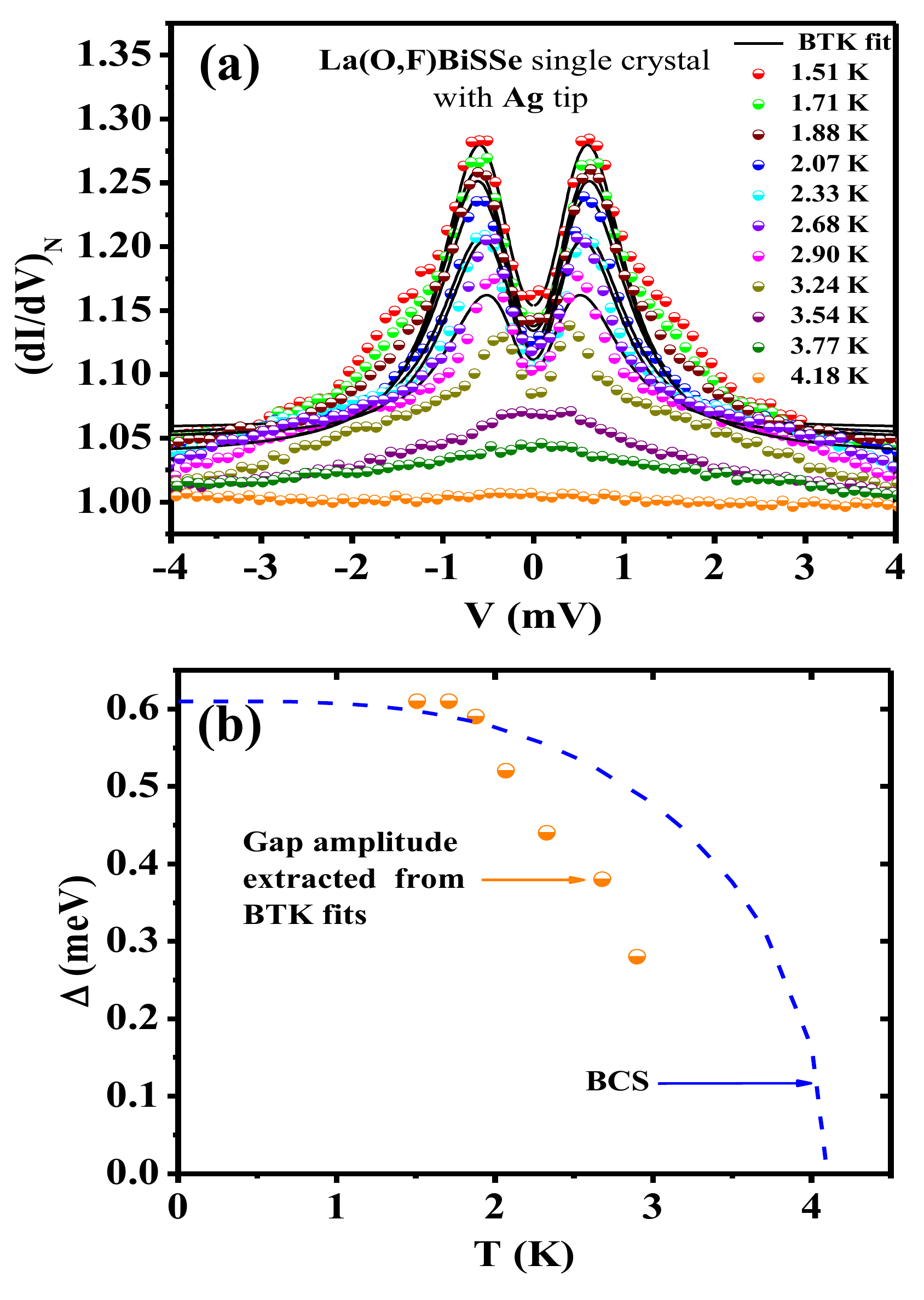}
	\caption{(\textbf{a}) Point contact spectra (dots)  with BTK fit (solid line) at different temperature below $T_c=4.2$ K obtained on La(O,F)BiSeS single crystal using silver (Ag) tip. Observed  double peak structure symmetric about V= 0  is a hallmark of Andreev reflection. The superconducting energy gap amplitude ($\Delta$) extracted by BTK fit is 0.61 meV. (\textbf{b}) The temperature dependence of superconducting gap  (orange dots) extracted from BTK fit do not follow conventional BCS behaviour.}
	\label{Figure 1}
\end{figure}

In Figure 1(a) we show temperature dependent PCAR spectrum between La(O,F)BiSeS and a metallic tip of Ag (dotted lines). Theoretical fits to the spectra using modified Blonder-Tinkham-Klapwijk (BTK) theory\cite{BTK} are also shown as solid lines. Modification to the BTK theory\cite{Placenick} was done to incorporate an inelastic broadening parameter ($\Gamma$) that takes care of broadening of the spectra. For lower temperature spectra, as it is seen, with a single gap BTK formula the lower bias part of the spectra fit well while a deviation is seen at higher bias. This deviation could be attributed to the existence of low-bias phonon modes \cite{P} or multiple superconducting gaps \cite{Souma,Tsuda,Mintu}. At relatively higher temperatures, the spectra fit very well with the modified BTK theory. 

In Figure 1(b) we show the temperature dependence of the superconducting energy gap as extracted from the data presented in Figure 1(a). The dotted  blue line in Figure 1(b) shows the expected temperature dependence for a conventional BCS superconductor. It is clearly seen that the experimentally measured temperature dependence deviates significantly from the BCS prediction\cite{tinkham}. Therefore, it can be concluded that the superconductivity in La(O,F)BiSeS is unconventional in nature. Such observation of an unconventional pairing is consistent with the recent theoretical prediction of the possibility of unconventional superconductivity in BiS$_2$-based superconductors.\cite{Baskaran}

\begin{figure}[h!]
	\centering
		\includegraphics[width=0.4\textwidth]{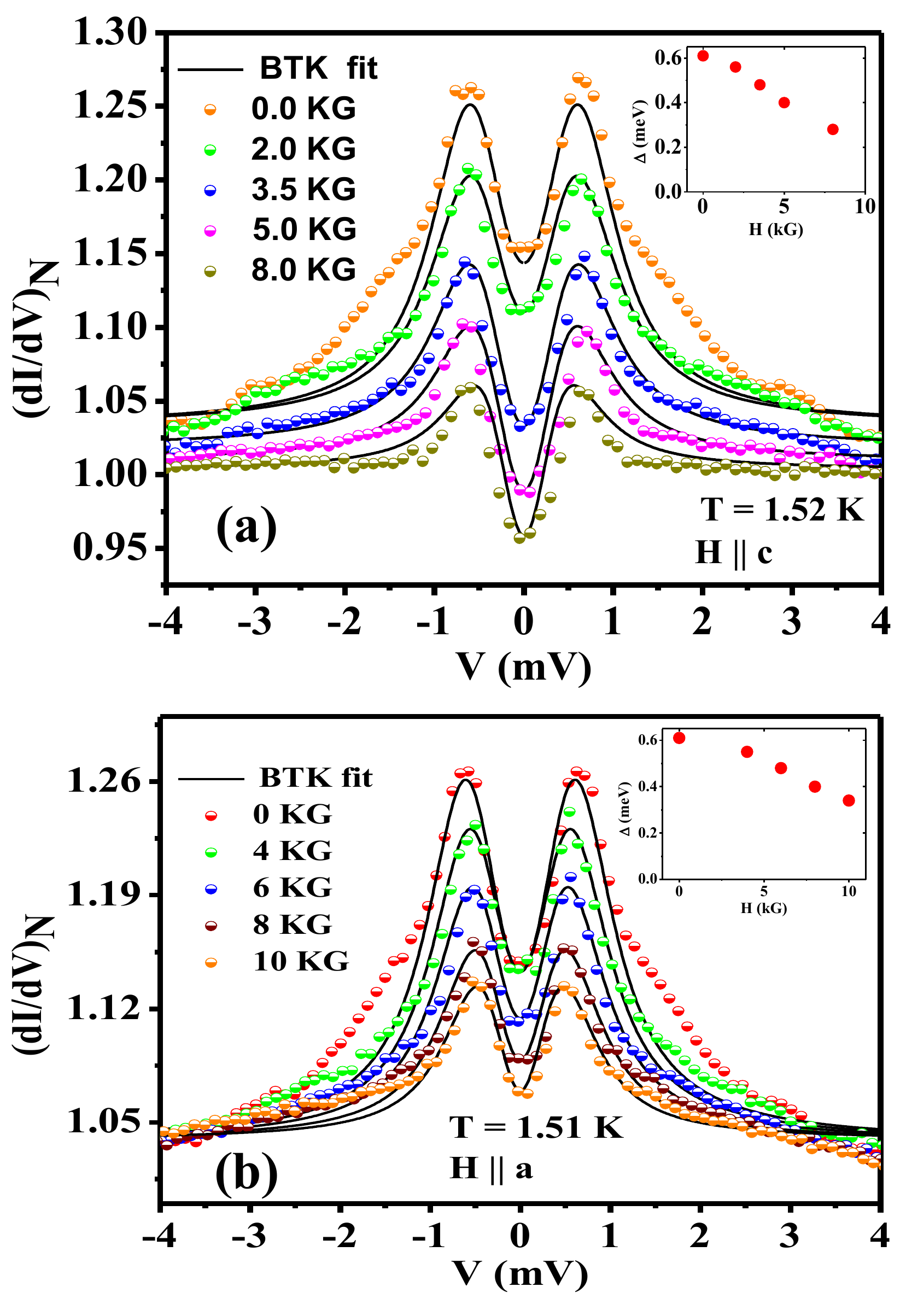}
	\caption{Magnetic field dependence of point contact spectra and corresponding superconducting energy gap extracted from BTK fit in the inset (\textbf{a}) along $c$-axis and (\textbf{b}) $a$-axis.}
	\label{Figure 2}
\end{figure}

In order to gain further understanding of the nature of superconductivity in La(O,F)BiSeS we have performed Andreev reflection spectroscopy at different magnetic fields applied along different crystallographic directions. Figure 2(a) shows the magnetic field dependence of the Andreev reflection spectra when the magnetic field is applied along the $c$-axis ($H||c$) of the crystal. The dotted lines show the experimental data points while the solid lines are the theoretical fits. In contrast to the temperature dependent measurements it is seen that at finite magnetic fields, the spectra fit very well with the modified BTK theory. This indicates that the deviation of the data from the fit seen in the zero-field spectrum might be due to the presence of additional superconducting gaps which get suppressed in small magnetic fields. The superconducting energy gap systematically decreases with increasing magnetic field. Though the gap is not clearly resolved, the features associated with superconductivity survive at higher fields up to 3 Tesla (in order to keep the panel uncluttered we have not shown the spectra above 8 kG) which can be considered to be the critical field of the superconducting point contact. The magnetic field dependence of the Andreev reflection spectra shows a different behaviour when the magnetic field is applied in the $ab$-plane ($H||a$) of the superconductor. The data for the $H||a$ is shown in Figure 2(b). The magnetic field dependence of the gaps for $H||c$ and $H||a$ are shown in the insets of Figures 2(a) and 2(b) respectively. It is seen that the amplitude of the superconducting energy gap decreases faster for $H||c$ than for $H||a$. At 8 kG, the measured gap for $H||c$ is 0.28 meV while that for $H||a$ is 0.4 meV exhibiting an anisotropy factor of 1.42 at 8 kG.

\begin{figure}[h!]
	\centering
		\includegraphics[width=0.4\textwidth]{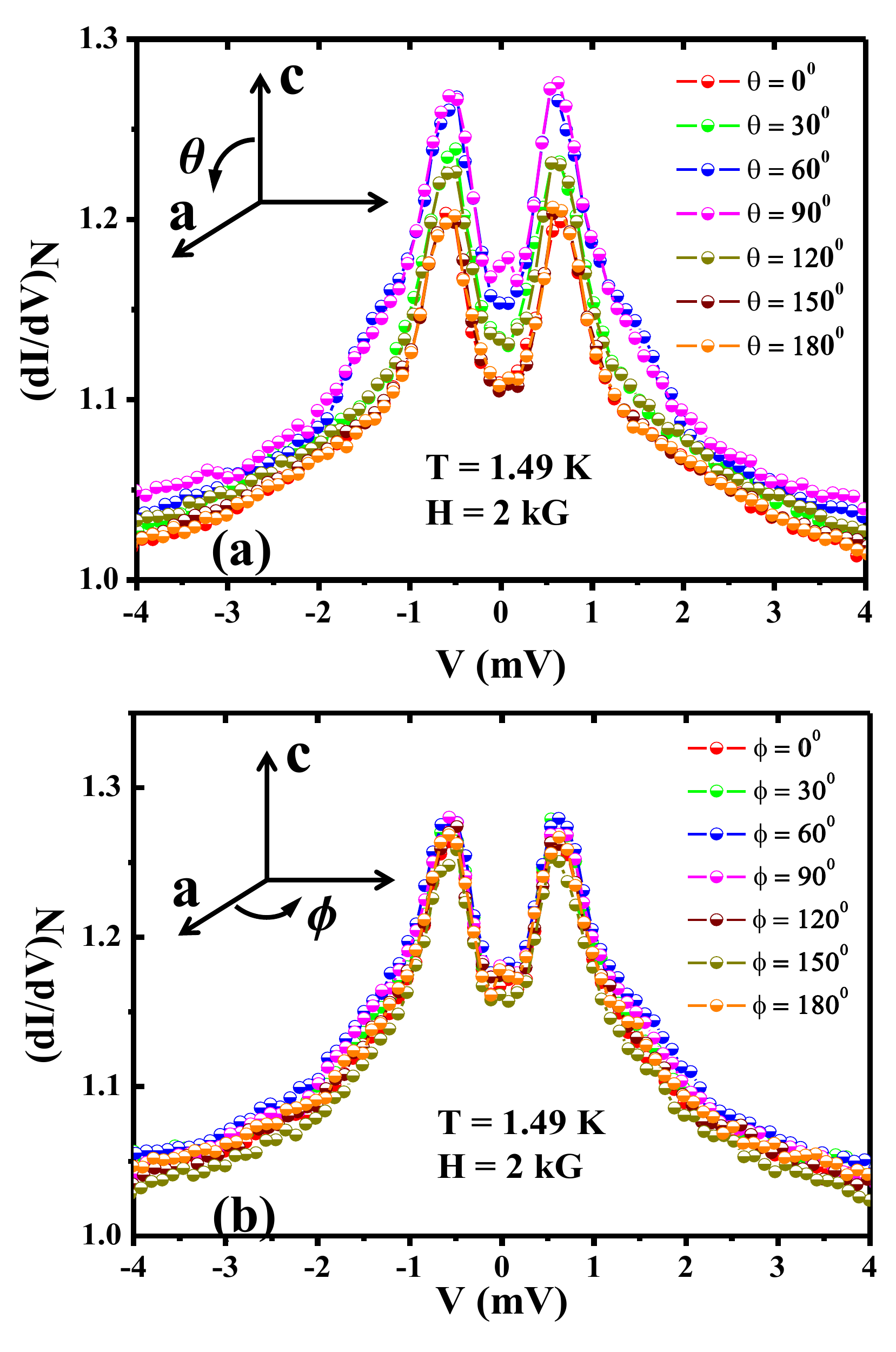}
	\caption{(\textbf{a}) Polar ($\phi=0$) and (\textbf{b}) azimuthal ($\theta=90^{\circ}$) angle dependence of point contact spectra at fixed magnetic  field of 2\,kG at $T=1.49$ K.}
	\label{Figure 3}
\end{figure}

From the discussion presented above it is clear that the superconductivity in La(O,F)BiSeS is unconventional in nature and there is a significant anisotropy\cite{Ume,Mukh} in the superconducting properties measured along the $c$-axis and along the $ab$-plane. Therefore, it is imperative to probe the anisotropic behaviour by changing the orientation ($\theta,\,\phi$) of the applied magnetic field continuously along different crystal axes. In order to probe that we have first applied a small magnetic field of 2 kG along the $c$-axis and then rotated the magnetic field with respect to the $c$-axis in the $ca$-plane. The data is shown in Figure 3(a). The spectrum evolves and the superconducting gap is seen to increase as the field direction changes from $H||c$ to $H||a$ and shows a maxima for $H||a$. This is consistent with the observed anisotropy in Figure 2(a) and Figure 2(b). In contrast, as shown in Figure 3(b), when the magnetic field angle ($\phi$) is changed in the $ab$-plane, the Andreev reflection spectra did not show noticeable change indicating that the anisotropy observed above is not observed in the $ab$-plane for a field of 2 kG. However, a similar anisotropy is observed for a field of 4 kG which is not shown here.

\begin{figure}[h!]
	\centering
		\includegraphics[width=0.5\textwidth]{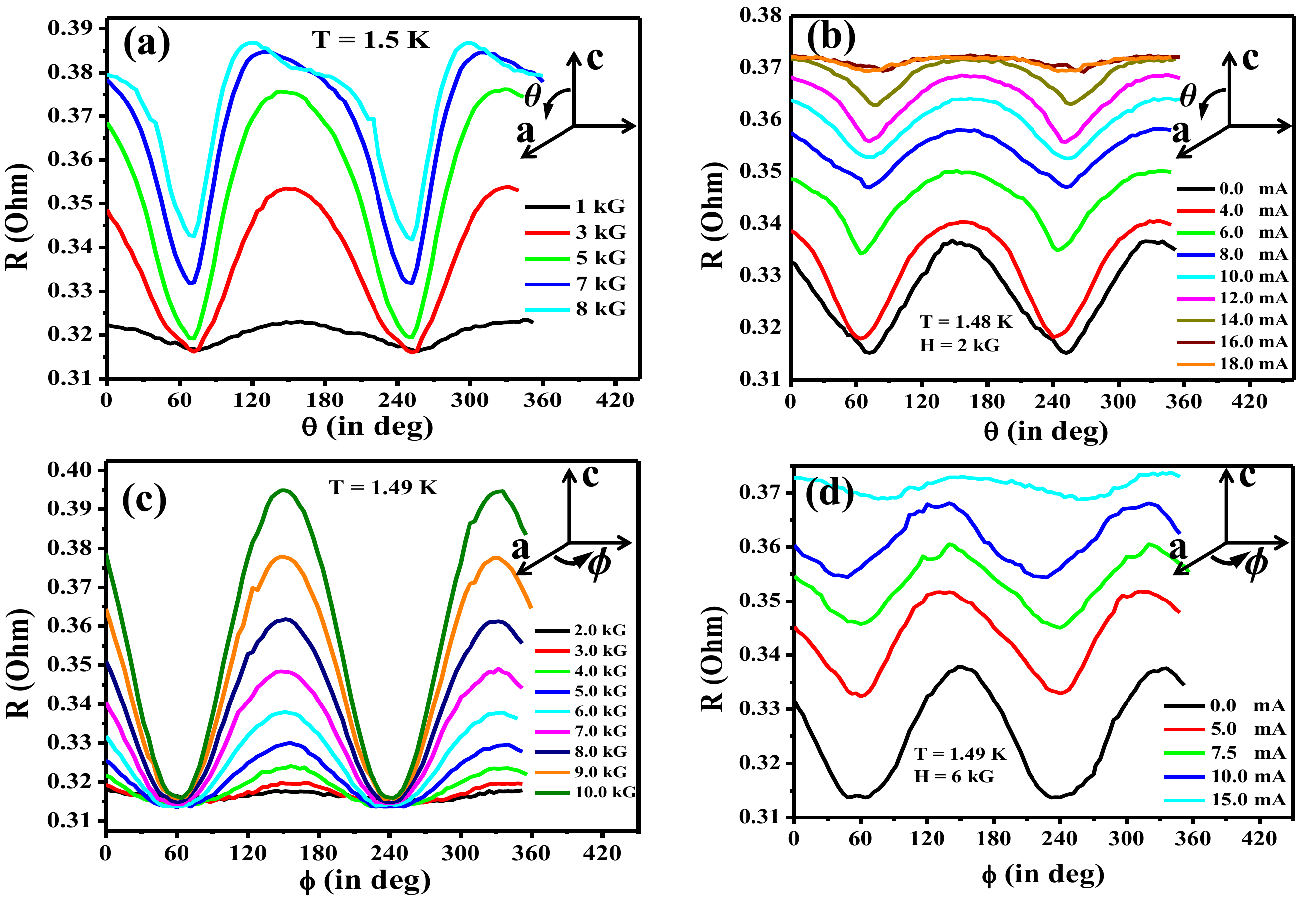}
	\caption{Polar magneto resistance (in $ca$-plane, $\phi=0$) for (\textbf{a}) different field strength and (\textbf{b}) different bias current.  Azimuthal magneto resistance (in $ab$-plane, $\theta=90^{\circ}$) for (\textbf{c}) different field strength and (\textbf{d}) different bias current.}
	\label{Figure 4}
\end{figure}
 
We have explored the anisotropic properties further by performing the field-angle dependence with increasing magnetic field along various directions and at different dc bias currents sent across the point contacts. In Figure 4(a) we show the $\theta$-dependence of the zero-bias resistance for different strengths of the applied magnetic field. During this measurement the azimuthal angle $\phi$ remained zero. It is clear that the anisotropy shows a two-fold symmetry\cite{V.P,Jin} for all values of the magnetic field and the anisotropy in zero-bias resistance increases with increasing the strength of the magnetic field. As shown in Figure 4(b), the anisotropy decreases as the dc current across the point contact is increased and the anisotropy completely disappears for a large current due to the critical current (of the given point contact) driven loss of superconductivity. This confirms that the anisotropy is directly associated with the superconducting phase of La(O,F)BiSeS. A similar two-fold symmetry is also observed (Figure 4(c)) when the field is rotated in the $ab$-plane for higher strengths of the magnetic field. In fact, for a field of 10 kG, the $ab$-plane anisotropy factor is almost 1.26. Even this anisotropy with $\phi$ disappears when the bias current across the point contact is increased far above the superconducting energy gap.

In order to discuss the observed anisotropy in the light of the physics of ballistic transport involving a normal metal and a superconductor with a momentum ($\vec{k}$)-dependent energy gap, it is first important to understand how the total current injected along a particular direction can be influenced by a magnetic field applied along another direction. For a given Fermi surface (with anisotropy in $\vec{k}$-space, if any) the current injected along a direction $\hat{n}$ is given by

\begin{equation*}
I\propto\oint\limits_{FS}N_{\vec{k}}(\vec{v}_{k}.\hat{n})dS_F={\langle N_{\vec{k}}\vec{v}_{k\hat{n}}\rangle}_{FS}
\end{equation*}
where $N_{\vec{k}}$ is  the  density  of  states, $\vec{v}_{k}.\hat{n}=\vec{v}_{k\hat{n}}$ is  the component of the Fermi velocity along $\hat{n}$ and $dS_F$ is an elementary area on the Fermi surface. For a non-spherical Fermi surface, the area of projection is different along different directions. In a point contact spectroscopy experiment with ballistic point contacts, the recorded spectra depends on the projection area of the Fermi surface on a plane perpendicular to $\hat{n}$, the direction of current flow. Therefore, when a superconducting energy gap $\Delta_{\vec{k}}$ with certain anisotropy is formed on the Fermi surface,  the average value of measured $\Delta$ for $I||\hat{n}$ is given by
 
\begin{equation*}
<\Delta>= {\langle\Delta_{\vec{k}}N_{\vec{k}}\vec{v}_{k\hat{n}}\rangle}_{FS}/{\langle N_{\vec{k}}\vec{v}_{k\hat{n}}\rangle}_{FS}.
\end{equation*}

When a magnetic field is applied along any $\vec{k}$-direction, the anisotropic superconducting gap $\Delta_{\vec{k}}$ for that direction will be suppressed slower or faster depending on the nature of the anisotropy of the upper critical field ($H_{c2}$), if any. Furthermore, the presence of an anisotropic energy gap itself might lead to an anisotropy in $H_{c2}$. As per the discussion above, such a change in $\Delta_{\vec{k}}$, for the field applied along $\vec{k}$ must affect the average $<\Delta>$ that is measured by injecting current along the $c$-axis as seen in Figure 2 and Figure 3. This idea is further supported by the fact that in Figure 3, when a field of 2 kG is rotated in the $ab$-plane, the Andreev reflection spectrum for $I||c$ remains unchanged. This is consistent because the effect of a small magnetic field applied along a perpendicular direction should be negligible.

Therefore, the idea of the existence of an anisotropic superconducting energy gap in La(O,F)BiSeS  explains the observed anisotropy of the measured gap for $I||c$ when the magnetic field is rotated along different directions in different planes. In addition, looking at the nature of the anisotropy in the point contact resistance measurements for different directions of the applied magnetic field, it can also be concluded that the superconducting gap in La(O,F)BiSeS has a complex symmetry that is two-fold in the $ca$-plane and two-fold in the $ab$-plane with different anisotropy factors in different planes. 
 
 
In conclusion, we have observed unconventional nature of superconductivity in La(O,F)BiSeS from temperature dependent Andreev reflection spectroscopy. From the field-angle dependence of the Andreev reflection spectra and the zero-bias resistance we observe a large anisotropy in the superconducting properties of La(O,F)BiSeS which indicates the existence of a two-fold symmetric upper critical field and an anisotrpic order paramater with complex symmetry in La(O,F)BiSeS.

We sincerely thank Dr. H. Takeya for his help. MA would like to thank CSIR, India for junior research fellowship (JRF). GS would like to thank Department of Science and Tenchnology (DST), Govt. of India for financial support through Ramanujan fellowship (grant number: SR/S2/RJN-99/2011) and a project under Nanomission (grant number:SR/NM/NS-1249/2013(G)). 
\\

\end{document}